\PassOptionsToPackage{table}{xcolor}
\documentclass[pdflatex,sn-mathphys-num,iicol]{sn-jnl}
\usepackage[T1]{fontenc}
\usepackage[utf8]{inputenc}
\usepackage{graphicx}
\usepackage{amsmath,amssymb}
\usepackage{subcaption}
\usepackage{etoolbox}
\usepackage{flafter}
\usepackage{stfloats}
\usepackage{placeins}
\setcitestyle{numbers,square}
\usepackage{tabularx,array}
\usepackage{xcolor}
\usepackage{bm}
\let\orcidlogo\relax
\usepackage{orcidlink}
\renewcommand{\orcid}[1]{\orcidlink{#1}}
\makeatletter
\newcommand{\authororcid}[1]{%
    \g@addto@macro\artauthors{\nobreak\,\orcid{#1}}%
}
\makeatother
\usepackage[title]{appendix}
\usepackage{tikz}
\usepackage{tikz-3dplot}
\usetikzlibrary{patterns,calc}
\graphicspath{{./}{pictures/}}
\newcommand{\figref}[1]{Figure~\ref{#1}}
\renewcommand{\eqref}[1]{Eq.~\ref{#1}}
\newcommand{\tabref}[1]{Table~\ref{#1}}
\definecolor{Blue}{HTML}{7F7FFF} 
\definecolor{Red}{HTML}{CD3031}
\definecolor{Green}{HTML}{20BF55}
\definecolor{Orange}{HTML}{FA9F42}
\definecolor{Aqua}{HTML}{00CEE0}
\definecolor{Pink}{HTML}{F8BDC4}
\definecolor{GrayLight}{HTML}{B3B3B3}
\definecolor{Gray}{HTML}{999999}

\begin{document}

\title{A graph neural network for deep learning-based full event interpretation at FCC-$ee$}

\author[1,2]{%
    \fnm{Elena}
    \sur{Graverini}%
}
\authororcid{0000-0003-4647-6429}

\author*[2]{%
    \fnm{Samuel}
    \sur{Morand}%
}
\authororcid{0009-0000-7031-9605}
\email{samuel.morand@cern.ch}

\author[3]{%
    \fnm{Rafael}
    \sur{Silva Coutinho}%
}
\authororcid{0000-0002-1545-959X}

\author[4]{%
    \fnm{Felipe Luan}
    \sur{Souza de Almeida}%
}
\authororcid{0000-0001-7181-6785}

\affil[1]{University of Pisa and INFN Pisa, Italy}
\affil[2]{École Polytechnique Fédérale de Lausanne (EPFL), Switzerland}
\affil[3]{Centro Brasileiro de Pesquisas Fisicas (CBPF), Brazil}
\affil[4]{Universitat de Barcelona, Spain}

\abstract{\unboldmath This paper presents a deep learning-based full event interpretation prototype developed for the electron--positron Future Circular Collider (FCC-$ee$) using simulated $e^{+}e^{-}\rightarrow Z\rightarrow b\bar{b}$ events at $\sqrt{s} = 91 \ \mathrm{GeV}$. The study was carried out using the Innovative Detector for Electron positron Accelerators (IDEA) detector concept as case study for future application. The proposed graph neural network framework aims to simultaneously identify and reconstruct the hierarchical decay trees of $b$-hadrons in each event using information from reconstructed charged particles. Events are represented as graphs, where reconstructed charged particles form the nodes and pairwise observables are encoded as edge features describing the relations between particles. On a test sample, 90\% of reconstructed decay chains contain exactly the correct set of particles, irrespective of their inferred decay hierarchy. Across a range of benchmark $b$-hadron decays, this clustering efficiency exceeds 93\% for decay chains spanning up to three generations of decays.}

\maketitle

\section{Introduction}

The Future Circular Collider (FCC) is CERN proposed post-LHC accelerator programme, conceived as a staged facility combining precision measurements and direct searches for New Physics (NP). Its first stage, FCC-$ee$, consists of a circular electron--positron collider with a circumference of approximately 90~km, foreseen to start operations in the mid-2040s. It will sequentially operate at four centre-of-mass energies: the $Z$ boson resonance ($\sqrt{s}=91.2$ GeV), the $W$ pair production threshold ($\sqrt{s}\approx 157.5$--$162.5$ GeV), the Higgsstrahlung maximum ($\sqrt{s}=240$ GeV) and the top quark pair production threshold ($\sqrt{s}\approx 340$--$365$ GeV). Over an expected operating period of about 15 years, FCC-$ee$ is projected to produce approximately $6\times10^{12}$ $Z$ bosons, $2.4\times10^{8}$ $W$ boson pairs, nearly $3\times10^{6}$ Higgs bosons and $2\times10^{6}$ $t\bar{t}$ pairs. Such unprecedented event samples will enable substantially more precise measurements of electroweak observables, Higgs couplings, top quark properties and flavour processes ~\cite{benedikt_fcc_2025-1}. Even if new particles are too massive to be produced directly, they may modify these observables through virtual corrections, giving FCC-$ee$ indirect sensitivity to NP scales reaching up to approximately 100 TeV. The second stage of the programme, FCC-hh, will reuse the same tunnel for $pp$ collisions at centre of mass energies approaching 100 TeV, extending the direct discovery reach far beyond that of the LHC~\cite{benedikt_fcc_2025}. The two stages are therefore complementary: FCC-$ee$ will probe new phenomena indirectly through precision measurements while FCC-hh will target their direct production. Several detector concepts are being studied for FCC-$ee$, including the Innovative Detector for Electron positron Accelerators (IDEA). With acceptance close to the whole \(4\pi\) solid angle, IDEA combines a vertex detector, a drift chamber and an outer silicon layer for precise tracking. Dual-readout electromagnetic and hadronic calorimeters provide accurate particle and jet energy measurements, while a 2 T solenoid and an outer muon system complete the detector~\cite{group_idea_2025}. This broad angular acceptance is advantageous for reconstructing complete event topologies. The FCC-$ee$ provides a cleaner collision environment than a hadron collider, thanks to a precisely known initial state and a low interaction rate. This rate is expected to be about $2\times10^{-3}$ per bunch crossing at the $Z$ pole and below $10^{-2}$ at the highest operating energies~\cite{benedikt_fcc_2025}. Together, the low interaction rate and broad detector coverage are expected to make FCC-$ee$ particularly favorable for flavor physics studies by enabling cleaner and more complete reconstruction of heavy hadron decays than in the $pp$ collision environment and forward detector geometry of LHCb, while providing substantially larger heavy flavor samples than B-factories such as Belle II.

Machine learning (ML), and deep learning in particular, has become increasingly important in particle physics, driven both by the rapid development of modern ML techniques and by the growing data volumes and event complexity associated with higher luminosity colliders. Recent developments include models such as ParticleNet, which represents jet constituents as particle clouds and applies dynamic graph convolutions to learn correlations between them. It is used to perform jet tagging and was subsequently adopted in CMS applications \cite{Qu:2019gqs}. At ATLAS, graph and transformer -based architectures such as GN1 and GN2 process track and secondary vertex information to identify jets containing heavy flavor hadrons~\cite{ATLAS:2022rkn,ATLAS:2025dkv}. At the level of detector hits, the Exa.TrkX programme applies graph neural networks to charged particle tracking by classifying candidate connections between hits under the high occupancy conditions expected at the High Luminosity LHC (HL-LHC)~\cite{ExaTrkX:2021abe}. The CMS Machine-Learned Particle Flow (MLPF) approach extends the use of neural networks to event reconstruction by processing tracks and calorimeter clusters and predicting the identities and four-momenta of reconstructed particles~\cite{Pata:2021oez,Pata:2023rhh}. These examples illustrate the application of deep neural networks at several stages of collider data processing, from detector hits and reconstructed particles to jets.

Beyond reconstructing the individual particles in an event, an additional challenge is to determine how they are relate within the underlying decay chains. Full Event Interpretation (FEI) was first developed by the Belle II collaboration for tag-side reconstruction in $e^{+}e^{-}\rightarrow\Upsilon(4S)\rightarrow B\bar{B}$ events, where reconstructing one $B$-meson provides kinematic and flavor constraints on the decay of the other one. The Belle II FEI algorithm uses a hierarchy of boosted decision trees trained on exclusive hadronic and semileptonic decay channels~\cite{keck_full_2019} and its reconstruction efficiency and signal coverage therefore remain tied to the modes explicitly implemented. A more general formulation of hierarchical decay tree reconstruction was introduced in~\cite{kahn_learning_2022} using Lowest Common Ancestor Generation (LCAG) labels. For each pair of final state particles, the label identifies the generation of their lowest common ancestor in the decay tree, reducing the reconstruction problem to pairwise classification. The same study explored several neural network architectures and showed that graph neural networks, following the general framework introduced in~\cite{battaglia_relational_2018}, outperform transformer-based approaches for this task. Building on this formulation, graph-based Full Event Interpretation (graFEI) was developed at Belle II to reconstruct generic $B$-meson decay structures~\cite{cerasoli_graph_2026}. Reconstructed final state particles are represented as graph nodes while the network predicts their hierarchical relationships and particle species. Graph-based methods have also been applied to flavor tagging. The Belle II Graph Flavor Tagger (GFlaT) uses charged particles associated with the accompanying $B$-meson to determine the production flavor of the signal-side neutral $B$-meson~\cite{adachi_new_2024}, while the LHCb Inclusive Flavor Tagger combines information from relevant tracks throughout $pp$ collisions to infer the initial flavor of reconstructed beauty hadrons~\cite{aaij_inclusive_2025}. Deep learning-based Full Event Interpretation (DFEI) was proposed at LHCb to identify, isolate and reconstruct the hierarchical decay chains of multiple heavy hadrons produced in the same $pp$ collision. Its key feature is its signal-agnostic approach: trained on inclusive $b\bar{b}$ samples, it can reconstruct even extremely rare decays absent from the training sample. DFEI also addresses the growing storage and online processing demands anticipated for future LHCb upgrades. By retaining only reconstructed objects associated with relevant heavy-flavor decays, it shifts the selection from choosing which complete events to store to deciding which parts of each event to preserve. The first prototype demonstrated the feasibility of reconstructing heavy-hadron decay trees from charged particles in simplified simulated events~\cite{pardinas_gnn_2023}. Subsequent work introduced heterogeneous graph neural networks (HGNN) to provide decay chain to $pp$ collision association and improve inference performance~\cite{sutcliffe_scalable_2025}, while further developments investigated the inclusion of neutral particles and increasingly realistic detector descriptions~\cite{bornand_deep-learning-based_2025}.

\begin{figure}[!htbp]
    \centering
    \includegraphics[
        width=0.95\linewidth,
        trim={6cm 2cm 4cm 0.1cm},
        clip
    ]{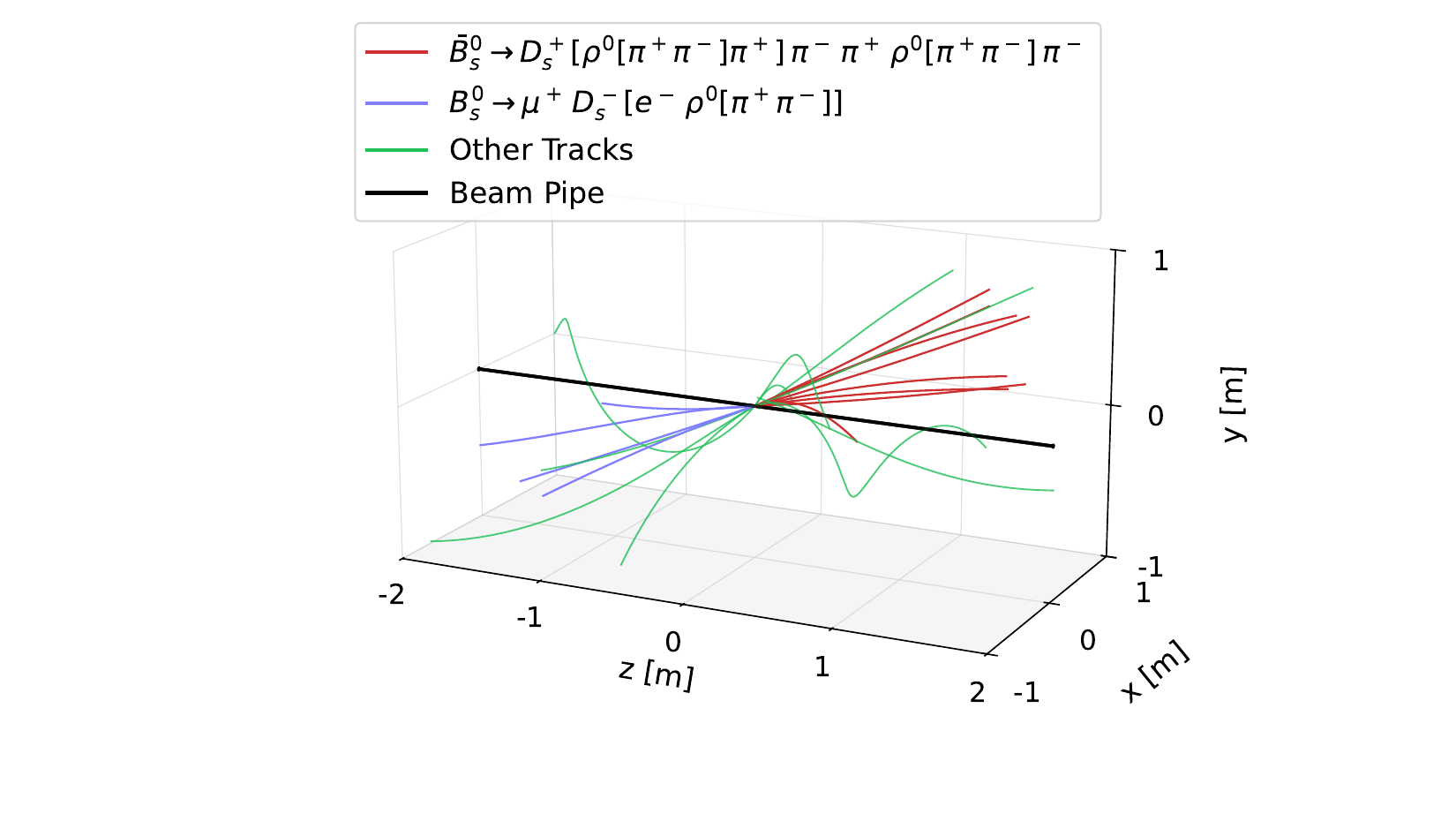}
    \caption{
    3D visualization of reconstructed tracks in a simulated
    \(e^{+}e^{-}\rightarrow Z\rightarrow b\bar{b}\) event.
    Tracks associated with the two $b$-hadron decay chains are shown
    in red and blue, while the remaining tracks are shown in green.
    The beam pipe is represented in black along the \(z\)-axis.
    }
    \label{fig:3d_helix}
\end{figure}

In this work, we investigate the performance of a DFEI-like approach at FCC-$ee$ using the IDEA detector concept as a case study. Reconstructed charged particles are represented as graph nodes and pairwise predictions are used to infer their hierarchical relationships within $b$-hadron decay chains without any predefined decay hypotheses, in a signal-agnostic way. This initial study considers charged particles only, and it focuses on hierarchical decay tree reconstruction without using any particle identification information. Future extensions will include neutral particle reconstruction, particle identification and additional detector information. The reconstruction task is illustrated in \figref{fig:3d_helix} for a simulated \(e^{+}e^{-}\rightarrow Z\rightarrow b\bar{b}\) event. The model must identify the charged particles belonging to each of the two $b$-hadron decay chains among the other particles in the event and reconstruct the decay relationships within each chain.

\section{Graph-based decay tree reconstruction}

\subsection{Dataset and graph representation}

The dataset consists of simulated inclusive
\(e^{+}e^{-}\rightarrow Z\rightarrow b\bar{b}\) events at
\(\sqrt{s}=91 \mathrm{GeV}\), produced within the FCC-$ee$ software  \cite{noauthor_hep-fccfccanalyses_2026}
using the IDEA detector configuration. Events are generated with
\texttt{Pythia8}~\cite{bierlich_comprehensive_2022}, heavy-hadron decays are
handled by \texttt{EvtGen}~\cite{lange_evtgen_2001} and detector response and
reconstruction are simulated with \texttt{Delphes}~\cite{favereau_delphes_2014}. The training and validation samples come from the Winter 2023 FCC simulation campaign \cite{noauthor_hep-fcceventproducer_2026}. Test samples, including benchmark $b$-physics channels, are produced locally using the same simulation framework. The simulated events are stored in EDM4hep format~\cite{carceller_edm4hep_2025} and converted to ROOT trees~\cite{brun_root_1997} for graph construction.

\begin{figure}[!htbp]
\centering
\resizebox{0.6\linewidth}{!}{%
\begin{tikzpicture}
\def\dzero{1.1}
\def\zzero{0.7}
\def\phizero{-55}
\def\radius{1.8}
\def\tanlambda{0.5}
\def\momlen{0.75}

\def\transDisplayRadius{1.0}

\def\arcAroundPOCA{35}
\pgfmathsetmacro{\helixStart}{-\arcAroundPOCA}
\pgfmathsetmacro{\helixEnd}{\arcAroundPOCA}
\def\helixSamples{60}

\def\transScale{1.55*1.347}
\def\longScale{1.55*1.347}

\def\mainBoxXmin{-1}
\def\mainBoxYmin{-1.35}
\def\mainBoxXmax{2.55}
\def\mainBoxYmax{1.75}

\def\transBoxXmin{-0.25}
\def\transBoxYmin{-0.25}
\def\transBoxXmax{2.35}
\def\transBoxYmax{2.05}

\def\longBoxXmin{-0.25}
\def\longBoxYmin{-0.25}
\def\longBoxXmax{2.35}
\def\longBoxYmax{2.05}

\pgfmathsetmacro{\mainBoxCX}{0.5*(\mainBoxXmin+\mainBoxXmax)}
\pgfmathsetmacro{\mainBoxCY}{0.5*(\mainBoxYmin+\mainBoxYmax)}

\pgfmathsetmacro{\transBoxCX}{0.5*(\transBoxXmin+\transBoxXmax)}
\pgfmathsetmacro{\transBoxCY}{0.5*(\transBoxYmin+\transBoxYmax)}

\pgfmathsetmacro{\longBoxCX}{0.5*(\longBoxXmin+\longBoxXmax)}
\pgfmathsetmacro{\longBoxCY}{0.5*(\longBoxYmin+\longBoxYmax)}

\def\sideOffset{5.6}

\def\transX{0}
\def\transY{\sideOffset}

\def\mainX{0}
\def\mainY{0}

\def\longX{0}
\def\longY{-\sideOffset}

\pgfmathsetmacro{\xzero}{-\dzero*sin(\phizero)}
\pgfmathsetmacro{\yzero}{\dzero*cos(\phizero)}

\node[inner sep=0pt] (main3d) at (\mainX,\mainY) {
\begin{tikzpicture}[scale=1.55]
\begin{scope}[shift={(-0.775,-0.2)}]

\draw[rounded corners, thick]
(\mainBoxXmin,\mainBoxYmin) rectangle (\mainBoxXmax,\mainBoxYmax);

\node[anchor=south west]
at (\mainBoxXmin + 1.25,\mainBoxYmax + 0.1)
{\small 3D View};

\tdplotsetmaincoords{70}{-60}

\begin{scope}[tdplot_main_coords]

\draw[->,black] (0,0,0) -- (2.5,0,0) node[right] {$z$};
\draw[->,black] (0,0,-0.8) -- (0,0,1.2) node[above] {$x$};
\draw[->,black] (0,0.6,0) -- (0,-1.2,0) node[right] {$y$};

\draw[black] (-1,0,0) -- (2.5,0,0);
\fill (0,0,0) circle (1.1pt);

\def\planesize{1.2}

\fill[pattern=north east lines, pattern color=Green, opacity=0.4]
(\zzero,-\planesize,-\planesize) --
(\zzero,\planesize,-\planesize) --
(\zzero,\planesize,\planesize) --
(\zzero,-\planesize,\planesize) -- cycle;

\draw[gray,dashed]
(\zzero,-\planesize,-\planesize) --
(\zzero,\planesize,-\planesize) --
(\zzero,\planesize,\planesize) --
(\zzero,-\planesize,\planesize) -- cycle;

\def\longzmin{\zzero}
\def\longzmax{2.1}

\fill[pattern=crosshatch dots, pattern color=Orange, opacity=0.55]
(\longzmin,-\planesize,-\planesize) --
(\longzmax,-\planesize,-\planesize) --
(\longzmax,-\planesize,\planesize) --
(\longzmin,-\planesize,\planesize) -- cycle;

\draw[gray,dashed]
(\longzmin,-\planesize,-\planesize) --
(\longzmax,-\planesize,-\planesize) --
(\longzmax,-\planesize,\planesize) --
(\longzmin,-\planesize,\planesize) -- cycle;

\fill[gray] (\zzero,0,0) circle (1.2pt);

\draw[gray,dashed]
(\zzero,-\planesize,0) -- (\zzero,\planesize,0);

\draw[gray,dashed]
(\zzero,0,-\planesize) -- (\zzero,0,\planesize);

\node[Gray]
at (\zzero,0,1.4)
{\small $z=z_0$};

\draw[
Blue,
very thick,
variable=\t,
samples=\helixSamples,
domain=\helixStart:\helixEnd
]
plot
(
{\zzero + \radius*\tanlambda*\t/57.2958},
{-\yzero - \radius*(sin(\phizero+\t)-sin(\phizero))},
{\xzero + \radius*(cos(\phizero+\t)-cos(\phizero))}
);

\fill[Red]
(\zzero,{-\yzero},\xzero)
circle (1.3pt);

\node[
Red,
left,
xshift=-0.15cm
]
at (\zzero,{-\yzero},\xzero)
{\small POCA};

\end{scope}
\end{scope}
\end{tikzpicture}
};

\node[inner sep=0pt] (transzoom) at (\transX,\transY) {
\begin{tikzpicture}[scale=\transScale]
\begin{scope}[shift={({-\transBoxCX},{-\transBoxCY})}]

\node[anchor=south west]
at (\transBoxXmin + 0.43,\transBoxYmax + 0.05)
{\small Transverse Projection};

\fill[pattern=north east lines, pattern color=Green, opacity=0.4]
(\transBoxXmin,\transBoxYmin)
rectangle
(\transBoxXmax,\transBoxYmax);

\draw[rounded corners,thick]
(\transBoxXmin,\transBoxYmin)
rectangle
(\transBoxXmax,\transBoxYmax);

\draw[->] (0,0) -- (1.75,0) node[right] {$y$};
\draw[->] (0,0) -- (0,1.45) node[above] {$x$};

\fill (0,0) circle (0.9pt);

\draw[
Blue,
very thick,
variable=\t,
samples=\helixSamples,
domain=\helixStart:\helixEnd
]
plot
(
{\yzero + \transDisplayRadius*(sin(\phizero+\t)-sin(\phizero))},
{\xzero + \transDisplayRadius*(cos(\phizero+\t)-cos(\phizero))}
);

\draw[-,thick,Red]
(0,0) -- (\yzero,\xzero)
node[midway,xshift=3pt,yshift=-13pt]
{$d_0$};

\draw[-,thick,Red]
(0,0.32)
arc[
start angle=90,
end angle={atan2(\xzero,\yzero)},
radius=0.32
];

\node[Red]
at (+0.15,0.50)
{$\phi_0$};

\pgfmathsetmacro{\ycenter}{
\yzero-\transDisplayRadius*sin(\phizero)
}

\pgfmathsetmacro{\xcenter}{
\xzero-\transDisplayRadius*cos(\phizero)
}

\draw[-,thick,Red]
(\ycenter,\xcenter) -- (\yzero,\xzero);

\node[Red,above right]
at (
{0.5*(\ycenter+\yzero)},
{0.5*(\xcenter+\xzero)+0.1}
)
{$\omega^{-1}$};

\fill[Blue]
(\ycenter,\xcenter)
circle (0.6pt);

\draw[->,Gray,thick]
(\yzero,\xzero) --
(
{\yzero+\momlen*cos(\phizero)},
{\xzero-\momlen*sin(\phizero)}
)
node[above right]
{$\vec p$};

\fill[Red]
(\yzero,\xzero)
circle (1.2pt);

\end{scope}
\end{tikzpicture}
};

\node[inner sep=0pt] (longzoom) at (\longX,\longY) {
\begin{tikzpicture}[scale=\longScale]
\begin{scope}[shift={({-\longBoxCX},{-\longBoxCY})}]

\node[anchor=south west]
at (\longBoxXmin + 0.4,\longBoxYmax + 0.05)
{\small Longitudinal Projection};

\fill[
pattern=crosshatch dots,
pattern color=Orange,
opacity=0.55
]
(\longBoxXmin,\longBoxYmin)
rectangle
(\longBoxXmax,\longBoxYmax);

\draw[rounded corners,thick]
(\longBoxXmin,\longBoxYmin)
rectangle
(\longBoxXmax,\longBoxYmax);

\draw[->] (0,0) -- (1.85,0) node[right] {$z$};
\draw[->] (0,0) -- (0,1.45) node[above] {$x$};

\fill (0,0) circle (0.9pt);

\draw[
Blue,
very thick,
variable=\t,
samples=\helixSamples,
domain=\helixStart:\helixEnd
]
plot
(
{\zzero + \radius*\tanlambda*\t/57.2958},
{\xzero + \radius*(cos(\phizero+\t)-cos(\phizero))}
);

\draw[black,dashed]
(\zzero,0) -- (\zzero,{\xzero+0.9});

\draw[-,thick,Red]
(0,\xzero) -- (\zzero,\xzero)
node[midway,above,yshift=0.1cm]
{$z_0$};

\draw[->,Gray,thick]
(\zzero,\xzero) --
(
{\zzero+\momlen*sin(35)},
{\xzero+\momlen*cos(35)}
)
node[above right]
{$\vec p$};

\pgfmathsetmacro{\momangle}{
atan2({+\momlen*cos(35)},{\momlen*sin(35)})
}

\draw[thick,Red]
(\zzero,{\xzero+0.48})
arc[
start angle=90,
end angle=\momangle,
radius=0.48
];

\node[Red]
at ({\zzero+0.17},{\xzero+0.6})
{$\lambda$};

\fill[Red]
(\zzero,\xzero)
circle (1.2pt);

\end{scope}
\end{tikzpicture}
};

\end{tikzpicture}
}
\caption{
Illustration of the helix track parametrization. The central panel shows the charged particle trajectory in three dimensions, while the transverse and longitudinal projections show the parameters \(d_0\), \(\phi_0\), \(\omega\), \(z_0\), and \(\lambda\), together with the point of closest approach (POCA).
}
\label{fig:track}
\end{figure}
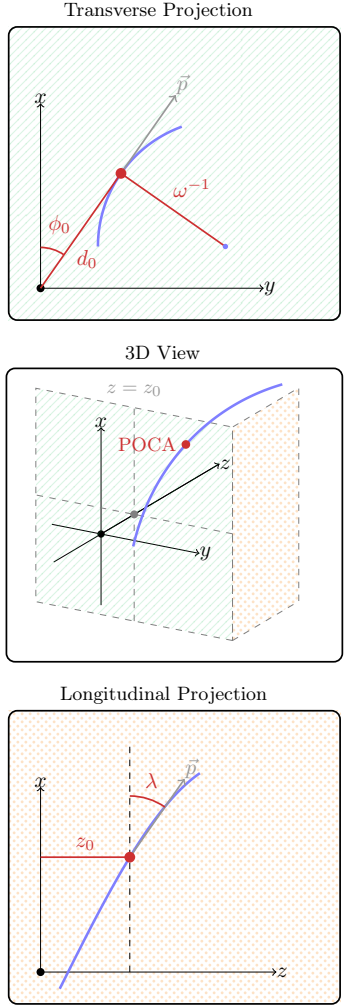

Consistent with the scope of this
study, only reconstructed particles associated with tracks are retained. Each event is represented as a fully connected undirected graph, where every reconstructed charged particle is connected to every other particle. The undirected structure reflects the symmetry of the pairwise particle relations. Each graph node contains kinematic and track information. The kinematic features are the momentum components \((p_x,p_y,p_z)\), the charge \(q\), the transverse momentum \(p_T\), and the pseudorapidity \(\eta\). Track information is provided by the helix parameters \(d_0, z_0, \phi_0, \tan\lambda, \omega\), whose definitions are illustrated in~\figref{fig:track}. The coordinates of the point of closest approach (POCA) are also included, together with the significance of each helix parameter, defined as the parameter value divided by its corresponding uncertainty.

\begin{figure}[!htbp]
\centering
\resizebox{0.8\linewidth}{!}{%
\begin{tikzpicture}[scale=1.7]

\tdplotsetmaincoords{70}{-58}

\begin{scope}[tdplot_main_coords]

\draw[->,black] (0,0,0) -- (3.0,0,0) node[right] {$z$};
\draw[->,black] (0,0,-0.5) -- (0,0,1.5) node[above] {$x$};
\draw[->,black] (0,1.0,0) -- (0,-1.4,0) node[right] {$y$};

\fill[black] (0,0,0) circle (1.1pt);

\def\dzeroA{1.35}
\def\zzeroA{0.25}
\def\phizeroA{-170}
\def\radiusA{1.6}
\def\tanlambdaA{0.9}

\def\dzeroB{0.95}
\def\zzeroB{1.85}
\def\phizeroB{-55}
\def\radiusB{1.25}
\def\tanlambdaB{-0.22}

\pgfmathsetmacro{\xA}{-\dzeroA*sin(\phizeroA)}
\pgfmathsetmacro{\yA}{ \dzeroA*cos(\phizeroA)}

\pgfmathsetmacro{\xB}{-\dzeroB*sin(\phizeroB)}
\pgfmathsetmacro{\yB}{ \dzeroB*cos(\phizeroB)}

\coordinate (POCAi) at (\zzeroA,{-\yA},\xA);
\coordinate (POCAj) at (\zzeroB,{-\yB},\xB);

\def\arcAroundPOCA{50}

\draw[
Blue,
very thick,
variable=\t,
samples=80,
domain=-\arcAroundPOCA:\arcAroundPOCA
]
plot
(
{\zzeroA + \radiusA*\tanlambdaA*\t/57.2958},
{-\yA - \radiusA*(sin(\phizeroA+\t)-sin(\phizeroA))},
{\xA + \radiusA*(cos(\phizeroA+\t)-cos(\phizeroA))}
);

\draw[
Red,
very thick,
variable=\t,
samples=80,
domain=-\arcAroundPOCA:\arcAroundPOCA
]
plot
(
{\zzeroB + \radiusB*\tanlambdaB*\t/57.2958},
{-\yB - \radiusB*(sin(\phizeroB+\t)-sin(\phizeroB))},
{\xB + \radiusB*(cos(\phizeroB+\t)-cos(\phizeroB))}
);

\pgfmathsetmacro{\normA}{sqrt(1+\tanlambdaA*\tanlambdaA)}
\pgfmathsetmacro{\uAz}{\tanlambdaA/\normA}
\pgfmathsetmacro{\uAy}{-cos(\phizeroA)/\normA}
\pgfmathsetmacro{\uAx}{-sin(\phizeroA)/\normA}

\pgfmathsetmacro{\normB}{sqrt(1+\tanlambdaB*\tanlambdaB)}
\pgfmathsetmacro{\uBz}{\tanlambdaB/\normB}
\pgfmathsetmacro{\uBy}{-cos(\phizeroB)/\normB}
\pgfmathsetmacro{\uBx}{-sin(\phizeroB)/\normB}

\def\tangentLength{0.75}

\draw[->,very thick,Blue]
(POCAi) --
({\zzeroA+\tangentLength*\uAz},
 {-\yA+\tangentLength*\uAy},
 {\xA+\tangentLength*\uAx})
node[above right] {$\vec{u}_i$};

\draw[->,very thick,Red]
(POCAj) --
({\zzeroB+\tangentLength*\uBz},
 {-\yB+\tangentLength*\uBy},
 {\xB+\tangentLength*\uBx})
node[above left] {$\vec{u}_j$};

\pgfmathsetmacro{\wz}{\zzeroA-\zzeroB}
\pgfmathsetmacro{\wy}{(-\yA)-(-\yB)}
\pgfmathsetmacro{\wx}{\xA-\xB}

\pgfmathsetmacro{\b}{\uAz*\uBz+\uAy*\uBy+\uAx*\uBx}
\pgfmathsetmacro{\d}{\uAz*\wz+\uAy*\wy+\uAx*\wx}
\pgfmathsetmacro{\e}{\uBz*\wz+\uBy*\wy+\uBx*\wx}

\pgfmathsetmacro{\denom}{1-\b*\b}

\pgfmathsetmacro{\sClosest}{(\b*\e-\d)/\denom}
\pgfmathsetmacro{\tClosest}{(\e-\b*\d)/\denom}

\pgfmathsetmacro{\docaIz}{\zzeroA+\sClosest*\uAz}
\pgfmathsetmacro{\docaIy}{-\yA+\sClosest*\uAy}
\pgfmathsetmacro{\docaIx}{\xA+\sClosest*\uAx}

\pgfmathsetmacro{\docaJz}{\zzeroB+\tClosest*\uBz}
\pgfmathsetmacro{\docaJy}{-\yB+\tClosest*\uBy}
\pgfmathsetmacro{\docaJx}{\xB+\tClosest*\uBx}

\def\lineBack{1.7}
\def\lineForward{1.7}

\draw[Blue,dashed,thick]
({\zzeroA-\lineBack*\uAz},
 {-\yA-\lineBack*\uAy},
 {\xA-\lineBack*\uAx})
--
({\zzeroA+\lineForward*\uAz},
 {-\yA+\lineForward*\uAy},
 {\xA+\lineForward*\uAx});

\draw[Red,dashed,thick]
({\zzeroB-\lineBack*\uBz},
 {-\yB-\lineBack*\uBy},
 {\xB-\lineBack*\uBx})
--
({\zzeroB+\lineForward*\uBz},
 {-\yB+\lineForward*\uBy},
 {\xB+\lineForward*\uBx});

\draw[Green,dashdotted,thick]
(\docaIz,\docaIy,\docaIx)
--
(\docaJz,\docaJy,\docaJx)
node[midway,above] {$\mathrm{DOCA}_{ij}$};

\fill[Green]
(\docaIz,\docaIy,\docaIx)
circle (0.8pt);

\fill[Green]
(\docaJz,\docaJy,\docaJx)
circle (0.8pt);

\fill[Orange] (POCAi) circle (1.4pt);
\fill[Orange] (POCAj) circle (1.4pt);

\node[Orange,left]
at (POCAi)
{\small POCA$_i$};

\node[Orange,right]
at (POCAj)
{\small POCA$_j$};

\end{scope}

\end{tikzpicture}
}

\caption{
Illustration of the approximation used to compute \(\mathrm{DOCA}_{ij}\).
The charged-particle trajectories are locally approximated by straight lines tangent to the tracks at their respective POCAs, and \(\mathrm{DOCA}_{ij}\) is obtained as the shortest distance between these lines.
}
\label{fig:doca_tangent}

\end{figure}
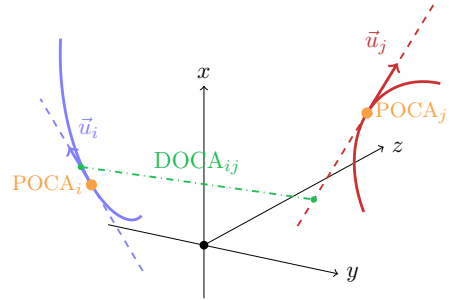

Edge features are defined for each pair of reconstructed particles \((i,j)\) and include the opening angle between their momenta \(\theta_{ij}\), the distance between their points of closest approach \(\|\Delta\vec{\mathrm{POCA}}_{ij}\|\) and its transverse and longitudinal components, the scalar product of the track directions at their POCAs \(\vec{u}_i\cdot\vec{u}_j\), the magnitudes of the total and transverse sums of their momenta \(\|\vec{p}_i+\vec{p}_j\|\) and \(\|\vec{p}_{T,i}+\vec{p}_{T,j}\|\), and the angle \(\alpha^{ij}\) between the pair’s total momentum and the average POCA of the two tracks. The distance of closest approach, \(\mathrm{DOCA}_{ij}\), is computed as the shortest distance between straight lines tangent to the tracks at their POCAs and included as an edge feature, as illustrated in~\figref{fig:doca_tangent}.

\begin{figure*}[!tp]
    \centering
    \includegraphics[width=0.95\linewidth]{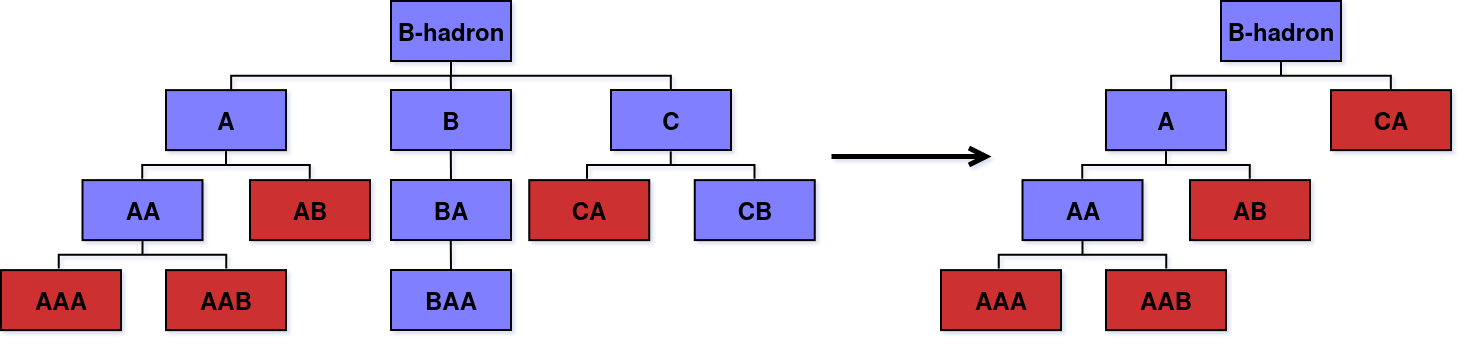}
    \caption{
    Example of the transformation from an ground truth decay tree (left) to its topologized representation (right). Reconstructed particles are shown in red and unreconstructed particles in blue. The topologized tree retains reconstructed leaves and branching points supported by at least two reconstructed branches.
    }
    \label{fig:topology}
\end{figure*}

The $b$-hadron decay trees used as training labels are identified from the ground truth relations, which describe the simulated decay trees rather than what the detector reconstructs. Ground-state $b$-hadrons serve as roots, and reconstructed particles are assigned to them by tracing their ancestry using the ground truth relations. The ground truth trees are then topologized following the approach introduced in Ref.~\cite{pardinas_gnn_2023}. Reconstructed particles are retained as leaves, while intermediate particles are kept only if at least two downstream branches contain reconstructed particles. Other intermediate states are removed and the remaining nodes reconnected, preserving the decay topology accessible from the reconstructed particles. In the example shown in~\figref{fig:topology}, particle A is kept because AB and either AAA or AAB are reconstructed, providing two downstream branches.

 The B branch is removed because none of its particles are reconstructed. Particle C is removed because only one of its downstream branches contains a reconstructed particle.

Only events containing a valid $b$-hadron decay tree and at most 50 reconstructed charged tracks are considered. The latter requirement removes a negligible fraction of the inclusive \(Z\rightarrow b\bar{b}\) sample. Events containing a topologized decay tree with depth greater than 3 are also discarded. This selection removes 12\% of the events from the sample. The selected event graphs are stored in Hierarchical Data Format version 5 (HDF5)~\cite{the_hdf_group_hierarchical_2026} and padded to 50 tracks to enable efficient data storage and loading.

\begin{figure}[!htbp]
    \centering
    \includegraphics[width=0.82\linewidth]{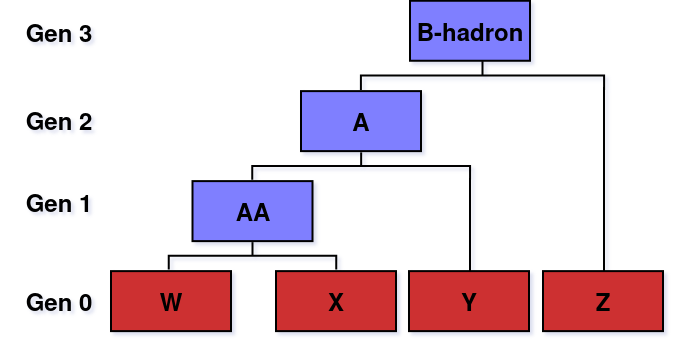}
\caption{
Example of a topologized decay tree with LCAG generations. Reconstructed particles are shown in red and correspond to graph nodes. In this example,
\(\mathrm{LCAG}(W,X)=1\),
\(\mathrm{LCAG}(X,Y)=\mathrm{LCAG}(W,Y)=2\) and
\(\mathrm{LCAG}(Y,Z)=3\).
}
    \label{fig:label}
\end{figure}

Once the graph representation is constructed, the corresponding edge labels are obtained from the topologized truth decay trees. The hierarchical decay structure is encoded using least common ancestor generation (LCAG) labels, following the formulation introduced in Ref.~\cite{kahn_learning_2022}. For each pair of reconstructed particles, the edge label corresponds to the LCAG in the topologized decay tree. A pull-down convention is used, in which reconstructed leaves have generation zero and successive ancestors are assigned increasing generations. If no valid common ancestor is found within the considered decay tree, the label is set to zero. An example is shown in~\figref{fig:label}.

\FloatBarrier
\subsection{Model architecture}

\begin{figure*}[!tp]
\centering
\includegraphics[width=0.70\linewidth]{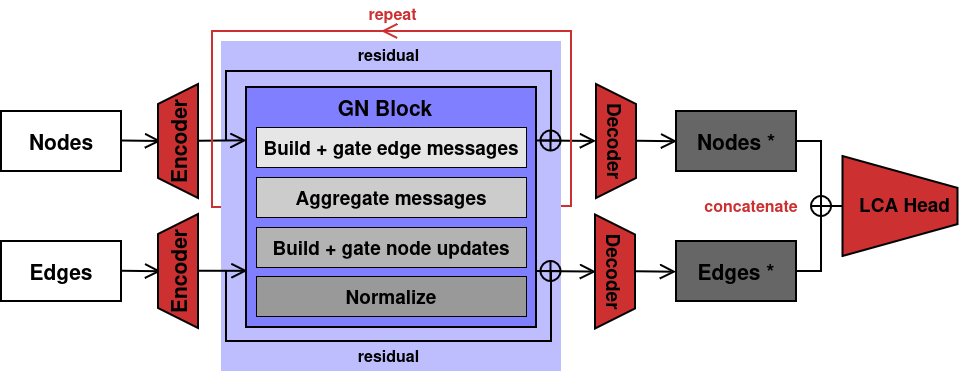}
\caption{
Schematic representation of the model architecture.
}
\label{fig:archi}
\end{figure*}

The model architecture is shown in~\figref{fig:archi}.

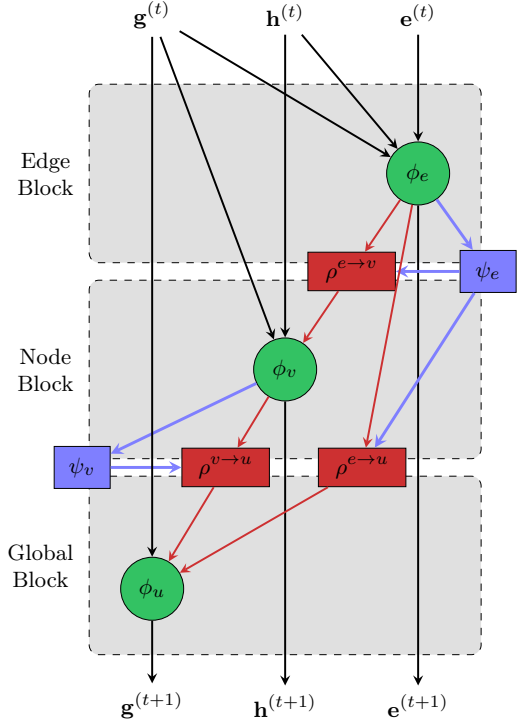
\begin{figure}[!htbp]
\centering
\resizebox{0.9\linewidth}{!}{%
\begin{tikzpicture}[
    >=stealth,
    block/.style={
        draw,
        dashed,
        rounded corners,
        fill=GrayLight,
        fill opacity=0.4,
        text opacity=1,
        minimum width=5.6cm,
        minimum height=2.55cm
    },
    op/.style={
        circle,
        draw,
        minimum size=0.9cm,
        font=\small,
        fill=Green,
        fill opacity=0.9,
        text opacity=1
    },
    gate/.style={
        rectangle,
        draw=black,
        fill=Blue,
        fill opacity=1,
        text opacity=1,
        minimum width=0.8cm,
        minimum height=0.6cm,
        font=\small
    },
    agg/.style={
        rectangle,
        draw=black,
        fill=Red,
        fill opacity=1,
        text opacity=1,
        minimum width=1.25cm,
        minimum height=0.55cm,
        font=\small
    },
    arrow/.style={->, thick},
    aggarrow/.style={->, thick, Red},
    bluearrow/.style={->, very thick, Blue},
    label/.style={font=\small}
]

\node[block] (edgeblock)   at (0,0) {};
\node[block] (nodeblock)   at (0,-2.8) {};
\node[block] (globalblock) at (0,-5.6) {};

\node[label, anchor=east, align=center] at (-2.9,0)
    {Edge\\Block};
\node[label, anchor=east, align=center] at (-2.9,-2.8)
    {Node\\Block};
\node[label, anchor=east, align=center] at (-2.9,-5.6)
    {Global\\Block};

\node[label] (u) at (-1.9,2.26667) {$\mathbf{g}^{(t)}$};
\node[label] (v) at (0,2.26667)    {$\mathbf{h}^{(t)}$};
\node[label] (e) at (1.9,2.26667)  {$\mathbf{e}^{(t)}$};

\node[label] (up) at (-1.9,-7.6) {$\mathbf{g}^{(t+1)}$};
\node[label] (vp) at (0,-7.6)    {$\mathbf{h}^{(t+1)}$};
\node[label] (ep) at (1.9,-7.6)  {$\mathbf{e}^{(t+1)}$};

\node[op]   (phie) at (1.9,0)    {$\phi_e$};
\node[gate] (psie) at (2.9,-1.4) {$\psi_e$};

\node[agg]  (rhoev) at (0.95,-1.4) {$\rho^{e\to v}$};
\node[op]   (phiv)  at (0,-2.8)    {$\phi_v$};
\node[gate] (psiv)  at (-2.9,-4.2) {$\psi_v$};

\node[agg] (rhoeu) at (1.1,-4.2)      {$\rho^{e\to u}$};
\node[agg] (rhovu) at (-0.85,-4.2)    {$\rho^{v\to u}$};
\node[op]  (phiu)  at (-1.9,-5.93333) {$\phi_u$};

\draw[arrow] (e) -- (phie);
\draw[arrow] (v) -- (phie);
\draw[arrow] (u) -- (phie);

\draw[arrow] (phie) -- (ep);

\draw[bluearrow] (phie) -- (psie);
\draw[bluearrow] (psie) -- (rhoev);
\draw[bluearrow] (psie) -- (rhoeu);

\draw[aggarrow] (phie) -- (rhoev);
\draw[aggarrow] (rhoev) -- (phiv);
\draw[arrow] (v) -- (phiv);
\draw[arrow] (u) -- (phiv);

\draw[bluearrow] (phiv) -- (psiv);
\draw[bluearrow] (psiv) -- (rhovu);

\draw[arrow] (phiv) -- (vp);

\draw[aggarrow] (phie) -- (rhoeu);
\draw[aggarrow] (phiv) -- (rhovu);

\draw[aggarrow] (rhoeu) -- (phiu);
\draw[aggarrow] (rhovu) -- (phiu);
\draw[arrow] (u) -- (phiu);

\draw[arrow] (phiu) -- (up);

\end{tikzpicture}%
}
\caption{%
Gated graph-network block used for weighted message passing.
Edge and node updates are modulated by learned gates before
information is propagated through the graph. The global branch
is shown for completeness. Adapted from
Refs.~\cite{battaglia_relational_2018,uzuki_improved_nodate}.
}
\label{fig:gn_block}
\end{figure}

Node and edge features are first projected independently into a common latent space using multilayer perceptrons (MLPs). The encoded graph is then processed by a sequence of graph-network blocks following the general framework of Ref.~\cite{battaglia_relational_2018}. A distinctive feature of the model is the use of learned gates during message passing, inspired by Ref.~\cite{uzuki_improved_nodate}. Since the graph is fully connected, not all particle-pair relations are expected to contribute equally to the decay tree reconstruction; the gates therefore regulate the information propagated through the graph and reduce the influence of less informative relations. Residual connections and layer normalization are applied between successive blocks. After the final block, the latent representations are decoded and the edge representations are passed to the LCAG classification head. The message-passing mechanism within each graph-network block is detailed in~\figref{fig:gn_block}.

The quantities $\mathbf{g}$, $\mathbf{h}$ and $\mathbf{e}$ denote the global, node and edge representations, while $\phi_e$, $\phi_v$ and $\phi_u$ are the corresponding update functions and $\rho^{e\to v}$, $\rho^{e\to u}$ and $\rho^{v\to u}$ the aggregation operators of the graph-network formalism~\cite{battaglia_relational_2018}. The edge update $\phi_e$ combines the current edge representation with symmetric combinations of the two connected node representations, ensuring invariance under their exchange. The resulting message is passed through the gate $\psi_e$, which produces a scalar weight between 0 and 1 and controls the edge contribution before the incident messages are aggregated at each node by summation through $\rho^{e\to v}$. The aggregated information is then combined with the current node representation through $\phi_v$, while a second gate, $\psi_v$, controls the magnitude of the node update. The functions $\phi_e$ and $\phi_v$ are implemented as MLPs, whereas $\psi_e$ and $\psi_v$ consist of a linear layer followed by a sigmoid activation. The global branch, involving $\rho^{e\to u}$, $\rho^{v\to u}$ and $\phi_u$, is shown for completeness but is not used in the present model.

\FloatBarrier
\subsection{Training}

The network is trained primarily to predict the LCAG label associated with each graph edge. The problem is formulated as a multiclass classification task, with the LCAG head yielding a class probability distribution. In this study, the LCAG labels are restricted to the classes 0, 1, 2 and 3, which are sufficient to describe all decay chain topologies with a depth of three or less, as motivated by the decay tree selection described above. The implementation can accommodate a larger number of LCAG classes. Since the LCAG distribution is strongly imbalanced, with class \(0\) dominating the training sample at a level of 90\%, class weights \(\alpha_k\) are introduced and chosen to be inversely proportional to the corresponding class frequencies. Two loss functions are considered for the LCAG objective. The first is weighted cross-entropy,
\begin{equation}
\mathcal{L}_{\mathrm{CE}}
=
-\sum_{k=1}^{K}
\alpha_k y_k \log(p_k) \ ,
\label{eq:weighted_ce}
\end{equation}
where \(y_k\) equals 1 if the true class is \(k\) and 0 otherwise, \(\alpha_k\) are the aforementioned weights, and \(p_k\) is the predicted probability for class \(k\). The second loss function is focal loss \cite{lin_focal_2018},
\begin{equation}
\mathcal{L}_{\mathrm{FL}}
=
-\sum_{k=1}^{K}
\alpha_k
(1-p_k)^{\gamma}
y_k \log(p_k) \ ,
\end{equation}
which reduces the contribution of examples assigned a high probability for their true class, focusing training on harder examples. The focal parameter \(\gamma\) controls the strength of this effect. In addition to the LCAG classification objective, the learned gates introduced in the message-passing blocks are constrained using further loss terms. These constraints are motivated by Ref.~\cite{uzuki_improved_nodate}, which showed that adding loss terms to constrain the gates improved model performance compared with leaving them unconstrained. For the node gates, the label is 1 for reconstructed particles originating from a $b$-hadron decay and 0 otherwise. For the edge gates, the label is 1 when the two connected particles belong to the same $b$-hadron decay tree, and 0 otherwise. The latter corresponds to edges with an LCAG label larger than 1. The node and edge gate outputs are compared with their respective binary labels using independent binary cross-entropy losses, denoted by \(\mathcal{L}_{\mathrm{node}}\) and \(\mathcal{L}_{\mathrm{edge}}\), of the same form as~\eqref{eq:weighted_ce}, but without class weights. Hence, the complete training objective is
\begin{equation}
\mathcal{L}
=
\mathcal{L}_{\mathrm{LCAG}}
+
\lambda_{\mathrm{node}}\mathcal{L}_{\mathrm{node}}
+
\lambda_{\mathrm{edge}}\mathcal{L}_{\mathrm{edge}} \ ,
\label{eq:total_loss}
\end{equation}
where \(\mathcal{L}_{\mathrm{LCAG}}\) denotes either the weighted cross-entropy or focal loss. The coefficients \(\lambda_{\mathrm{node}}\) and \(\lambda_{\mathrm{edge}}\) control the relative contribution of the additional loss terms and are treated as hyperparameters.

The training set consists of 90\% of the dataset, which contains 1.6 million events, selected randomly, while the remaining 10\% is used for validation. Training is performed with a batch size of 64. A cosine annealing learning rate schedule is used during training~\cite{loshchilov_sgdr_2017}, with the learning rate decreasing from \(10^{-3}\) to \(10^{-5}\). The cosine period is set equal to the maximum number of training epochs, which is 200. Early stopping is applied with a patience of 10 epochs, requiring an improvement of at least \(10^{-4}\) in the validation balanced accuracy. Hyperparameter optimization is performed with Optuna~\cite{akiba_optuna_2019}, using the balanced accuracy over the LCAG classes on the validation set as the performance metric. Since this metric exhibits epoch-to-epoch fluctuations, an exponential moving average is used to provide a more stable estimate of the validation performance. This avoids the additional computational cost of evaluating each configuration over multiple random seeds.

\subsection{From graphs to decay trees}

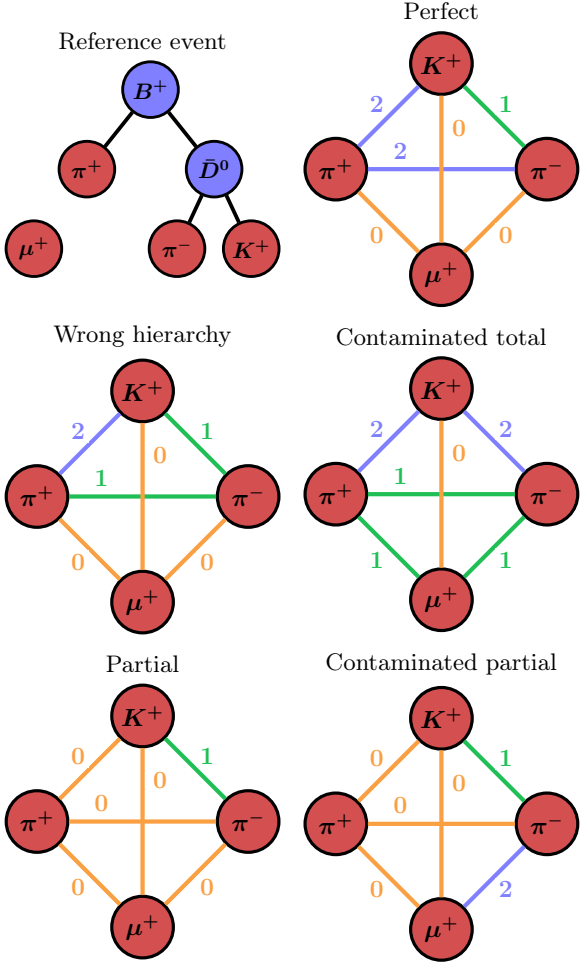
\begin{figure}[!htbp]
\centering

\newcommand{\edgestyle}[1]{%
  \ifnum#1=0 edgezero\else
  \ifnum#1=1 edgeone\else
  edgetwo\fi\fi
}

\newcommand{\decaytopology}[6]{%
\begin{tikzpicture}[
    scale=0.78,
    transform shape,
    particle/.style={
        circle,
        draw=black,
        fill=Red,
        fill opacity=0.85,
        text opacity=1,
        line width=1.2pt,
        minimum size=1.05cm,
        inner sep=0pt,
        font=\large
    },
    edgezero/.style={line width=1.6pt, Orange},
    edgeone/.style={line width=1.6pt, Green},
    edgetwo/.style={line width=1.6pt, Blue},
    edgelabel/.style={
        fill=white,
        inner sep=1.5pt,
        font=\large
    }
]
\node[particle] (K)   at (0, 1.8) {$\bm{K^{+}}$};
\node[particle] (pim) at (-1.8, 0) {$\bm{\pi^{+}}$};
\node[particle] (pip) at (1.8, 0) {$\bm{\pi^{-}}$};
\node[particle] (mu)  at (0, -1.8) {$\bm{\mu^{+}}$};

\draw[\edgestyle{#1}] (K) -- node[edgelabel, above left] {$\boldsymbol{#1}$} (pim);
\draw[\edgestyle{#2}] (K) -- node[edgelabel, above right] {$\boldsymbol{#2}$} (pip);
\draw[\edgestyle{#3}] (pim) -- node[edgelabel, above, yshift=0.1cm, xshift=-0.7cm] {$\boldsymbol{#3}$} (pip);
\draw[\edgestyle{#4}] (pim) -- node[edgelabel, below left] {$\boldsymbol{#4}$} (mu);
\draw[\edgestyle{#5}] (pip) -- node[edgelabel, below right] {$\boldsymbol{#5}$} (mu);
\draw[\edgestyle{#6}] (K) -- node[edgelabel, right, yshift=0.7cm, xshift=0.1cm] {$\boldsymbol{#6}$} (mu);
\end{tikzpicture}%
}

\newcommand{\topologywithtitle}[7]{%
\begin{minipage}{0.48\linewidth}
    \centering
    {\small #1\par}
    \vspace{0.12cm}
    \resizebox{\linewidth}{!}{\decaytopology{#2}{#3}{#4}{#5}{#6}{#7}}
\end{minipage}%
}

\newcommand{\truedecaytree}{%
\begin{tikzpicture}[
    scale=0.78,
    transform shape,
    particle/.style={
        circle,
        draw=black,
        fill=Red,
        fill opacity=0.85,
        text opacity=1,
        line width=1.2pt,
        minimum size=1.05cm,
        inner sep=0pt,
        font=\large
    },
    blueparticle/.style={
        circle,
        draw=black,
        fill=Blue,
        line width=1.2pt,
        minimum size=1.05cm,
        inner sep=0pt,
        font=\large
    },
    edge/.style={
        line width=1.6pt,
        black
    }
]
\node[blueparticle] (B)   at (0, 1.5) {$\bm{B^{+}}$};
\node[blueparticle] (D)   at (1.2, 0) {$\bm{\bar{D}^{0}}$};
\node[particle]     (piB) at (-1.2, 0) {$\bm{\pi^{+}}$};

\node[particle] (K)   at (1.9, -1.5) {$\bm{K^{+}}$};
\node[particle] (piD) at (0.5, -1.5) {$\bm{\pi^{-}}$};
\node[particle] (mu)  at (-2.2, -1.5) {$\bm{\mu^{+}}$};

\draw[edge] (B) -- (D);
\draw[edge] (B) -- (piB);
\draw[edge] (D) -- (K);
\draw[edge] (D) -- (piD);
\end{tikzpicture}%
}

\newcommand{\truedecaywithtitle}{%
\begin{minipage}{0.48\linewidth}
    \centering
    {\small Reference event\par}
    \vspace{0.12cm}
    \resizebox{\linewidth}{!}{\truedecaytree}
\end{minipage}%
}

\truedecaywithtitle
\hfill
\topologywithtitle{Perfect}{2}{1}{2}{0}{0}{0}

\medskip

\topologywithtitle{Wrong hierarchy}{2}{1}{1}{0}{0}{0}
\hfill
\topologywithtitle{Contaminated total}{2}{2}{1}{1}{1}{0}

\medskip

\topologywithtitle{Partial}{0}{1}{0}{0}{0}{0}
\hfill
\topologywithtitle{Contaminated partial}{0}{1}{0}{0}{2}{0}

\caption{
Illustration of the reconstruction categories with respect to a reference event containing a $B^+\to\bar{D}^0\pi^+$ decay. The event also contains an isolated muon, produced by an additional decay tree not shown. Red nodes represent reconstructed particles, while the edge labels correspond to the predicted LCAG classes.
}
\label{fig:physics_eval}
\end{figure}

For each pair of reconstructed charged particles \((i,j)\), the model predicts a probability distribution over the LCAG classes. The probability that the two particles belong to the same decay tree is defined as \(
p_{\mathrm{same}}^{ij}
=
1-p_0^{ij} \)
where \(p_0^{ij}\) is the predicted probability for LCAG class \(0\). Two particles are considered connected if \(
p_{\mathrm{same}}^{ij}
>
t_{\mathrm{conn}} \),
where \(t_{\mathrm{conn}}\) is a connectivity threshold, treated as a hyperparameter. The resulting connections define different clusters, which are interpreted as reconstructed decay chain candidates. This procedure does not require all pairwise LCAG predictions within a decay tree to be correct, provided that enough valid connections remain to preserve the correct set of reconstructed charged particles within the cluster. Within each cluster, the non-zero LCAG predictions are used to reconstruct the internal hierarchy. The same procedure is applied to the LCAG labels to obtain the reference ground truth decay trees. The reconstruction quality is evaluated by comparing each reconstructed candidate with the corresponding reference decay tree. Five exclusive reconstruction categories are defined:
\begin{itemize}
    \item \textbf{Perfect}: all particles in the decay tree are correctly inserted into the cluster, with the correct hierarchy;
    \item \textbf{Wrong hierarchy}: all particles are correctly inserted into the cluster, but at least one hierarchical relation is incorrect;
    \item \textbf{Contaminated total}: all true particles are correctly inserted into the cluster, together with additional particles not belonging to the true decay chain;
    \item \textbf{Partial}: only a subset of the true decay chain is recovered;
    \item \textbf{Contaminated partial}: only a subset of the true decay chain is recovered, together with additional particles not belonging to the true decay chain.
\end{itemize}
A schematic illustration of these categories is shown in~\figref{fig:physics_eval}.

The connectivity threshold is optimized on the validation set by maximizing the figure of merit defined by,
\begin{equation}
    \mathrm{FoM}=1.5\,f_{\mathrm{perfect}}+f_{\mathrm{WH}}\ ,
    \label{eq:fom}
\end{equation} 
where \(f_{\mathrm{perfect}}\) and \(f_{\mathrm{WH}}\) are the fractions of perfect and wrong hierarchy reconstructions, respectively. The larger weight assigned to the perfect category reflects the preference for fully correct reconstruction of both the particle content and the decay hierarchy. Since both categories recover the correct particle content, the clustering efficiency is defined as \(\epsilon = f_{\mathrm{perfect}} + f_{\mathrm{WH}}\). This metric directly measures the ability of the model to identify the correct set of particles associated with a decay chain, independently of errors in the reconstructed internal hierarchy.

\section{Results}

The hyperparameter optimization procedure yields a model with seven graph-network blocks and a focal loss characterized by  \(\gamma=1\). The auxiliary loss coefficients are found to be \(\lambda_{\mathrm{node}}=0.02\) and \(\lambda_{\mathrm{edge}}=0.23\). The optimization of these two coefficients is shown in Appendix~\ref{app:hyperparameter_optimization}. These values are used for all results presented below. 

\begin{figure}[!htbp]
    \centering
    \includegraphics[width=0.95\linewidth]{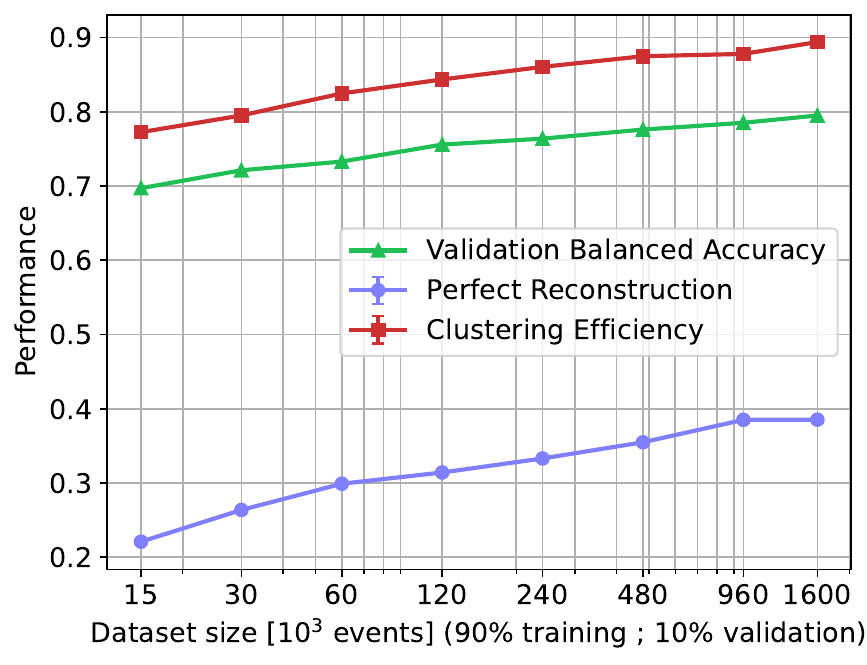}
    \caption{Model performance as a function of the training dataset size. The validation balanced accuracy, perfect reconstruction fraction and clustering efficiency are shown.}
    \label{fig:perf_size}
\end{figure}

First, the dependence of the model performance on the size of the training dataset is investigated. The validation balanced accuracy, perfect reconstruction fraction and clustering efficiency are shown in~\figref{fig:perf_size} as functions of the number of events in the training dataset.

 All three metrics improve as the dataset size increases. No clear saturation is observed within the considered range, suggesting that further gains may still be obtained with larger training samples. The final model used for the results presented below is trained on the largest available training dataset, containing \(1.6\times10^{6}\) events.

\begin{figure}[!htbp]
    \centering
    \includegraphics[width=0.95\linewidth]{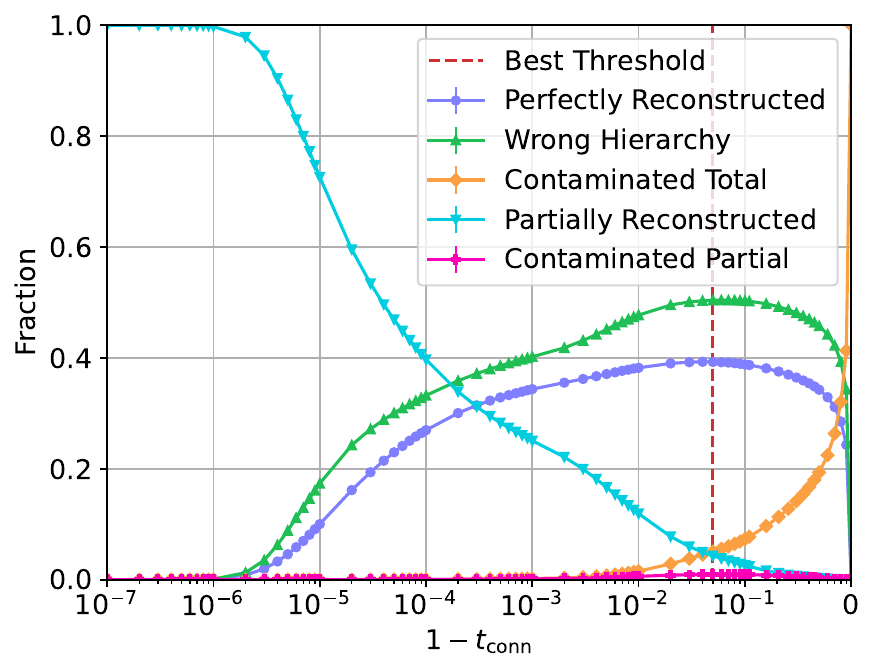}
    \caption{Fractions of the different reconstruction categories as a function of \(1-t_{\mathrm{conn}}\). The selected working point is indicated in red.}
    \label{fig:fom_cat}
\end{figure}

After training, the connectivity threshold \(t_{\mathrm{conn}}\) is optimized on the validation set by maximizing the figure of merit defined in \eqref{eq:fom}. The evolution of the reconstruction categories as a function of the threshold is shown in~\figref{fig:fom_cat}.

 Small values of \(t_{\mathrm{conn}}\) lead to more permissive clustering and therefore increase the fraction of contaminated total reconstructions. Larger values reduce contamination but increase the fraction of partial reconstructions. The selected working point maximizes the figure of merit, thus providing the best balance between these two effects.

\subsection{Performance on inclusive decays}\label{sec:perf_incl}

The performance on an inclusive \(e^{+}e^{-}\rightarrow Z\rightarrow b\bar{b}\) test sample containing 10\,000 events is summarized in \tabref{tab:incl_res}. The model achieves a perfect reconstruction fraction of 39\%, while 51\% of the decay trees are reconstructed with the correct particle content but an incorrect internal hierarchy. The resulting clustering efficiency is 89\%. This clustering efficiency is evaluated for clusters, while the corresponding balanced accuracy of \((79.75 \pm 0.07)\%\) is evaluated for individual edges connecting particle pairs. The remaining reconstruction categories contribute significantly less, with contaminated total and partial reconstructions accounting for 5\% and 4\%, respectively, while contaminated partial reconstructions represent less than 1\% of the inferences.
\begin{table}[tbp]
\caption{
Performance on the inclusive \(e^{+}e^{-}\rightarrow Z\rightarrow b\bar{b}\) test sample. Reconstruction categories are expressed as fractions of reconstructed decay trees.
}
\label{tab:incl_res}
\centering
\renewcommand{\tabularxcolumn}[1]{m{#1}}
\begin{tabularx}{\linewidth}{|X|>{\centering\arraybackslash}m{3cm}|}
\hline
\rowcolor{GrayLight!60}
\textbf{Metric} & \textbf{Value} \\
\hline
Clustering efficiency \newline (Perfect + Wrong hierarchy) &
$(89.4 \pm 0.2)$\% \\
\hline

\rowcolor{GrayLight!60}
Perfect & $(38.7 \pm 0.4)$\% \\
\hline

Wrong hierarchy & $(50.6 \pm 0.4)$\% \\
\hline

\rowcolor{GrayLight!60}
Contaminated total & $(5.3 \pm 0.2)$\% \\
\hline

Partial & $(4.4 \pm 0.2)$\% \\
\hline

\rowcolor{GrayLight!60}
Contaminated partial & $(0.9 \pm 0.1)$\% \\
\hline
\end{tabularx}
\end{table}

\begin{figure}[!htbp]
    \centering
    \includegraphics[width=0.95\linewidth]{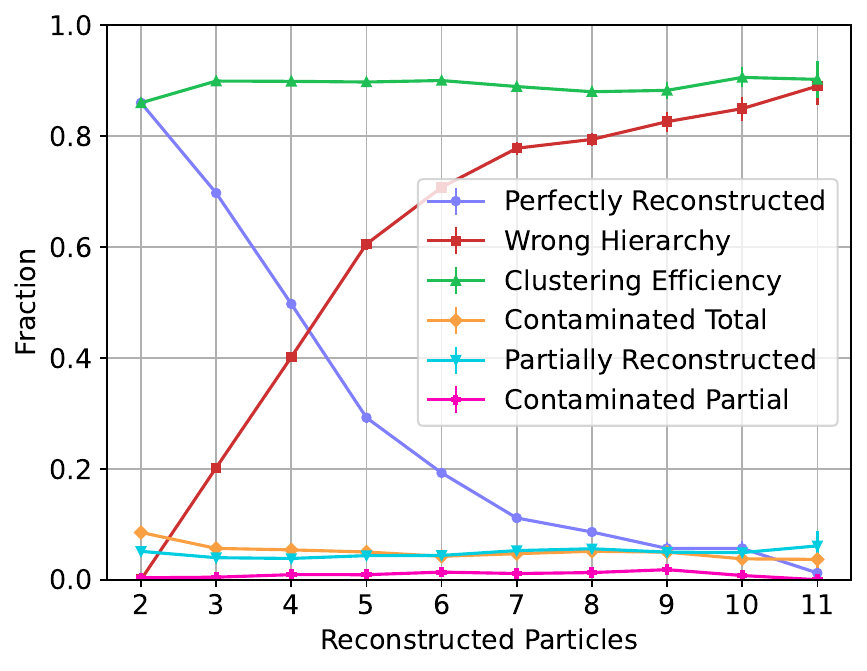}
    \caption{
    Fractions of the different reconstruction categories as a function of the number of reconstructed particles in the true decay tree, evaluated on the inclusive \(e^{+}e^{-}\rightarrow Z\rightarrow b\bar{b}\) test sample.
    }
    \label{fig:reco_frac}
\end{figure}

The reconstruction performance is further studied as a function of the multiplicity of reconstructed charged particles in the true decay tree. The fractions of the different reconstruction categories are shown in \figref{fig:reco_frac}.

As the particle multiplicity increases, the perfect reconstruction fraction decreases while the wrong hierarchy fraction increases. The sum of these two categories, the clustering efficiency, remains approximately stable over a broad range of multiplicities, indicating that the main difficulty for larger decay trees is the reconstruction of the internal hierarchy rather than the identification of the correct particle content.

\begin{figure}[!htbp]
    \centering
    \begin{subfigure}{0.98\linewidth}
        \centering
        \includegraphics[width=\linewidth]{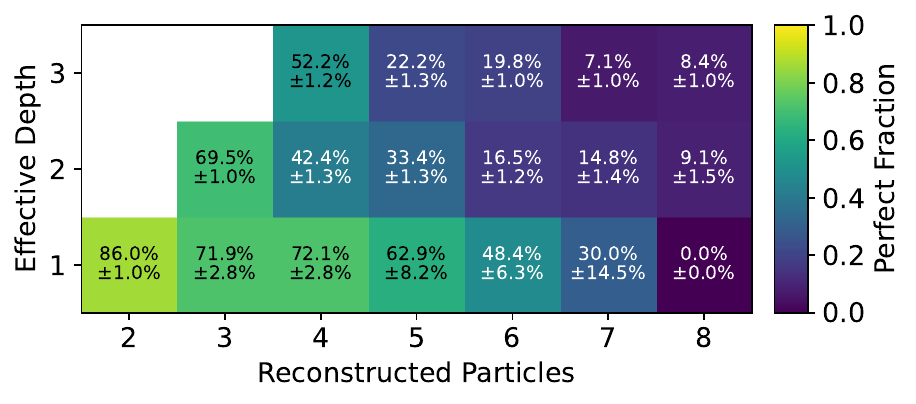}
        \caption{}
        \label{fig:heatmap_perf}
    \end{subfigure}

    \par\medskip

    \begin{subfigure}{0.98\linewidth}
        \centering
        \includegraphics[width=\linewidth]{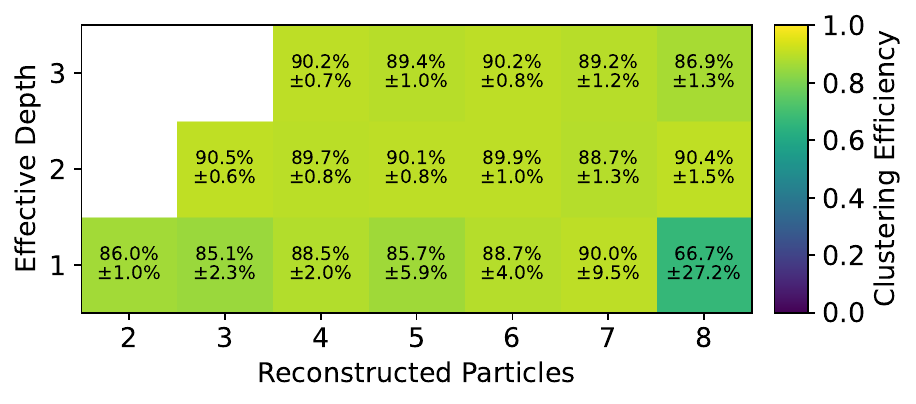}
        \caption{}
        \label{fig:heatmap_perf_wh}
    \end{subfigure}
    \caption{
    (a) Perfect reconstruction fraction and
    (b) clustering efficiency as functions of the topological decay tree depth and the number of reconstructed particles in the true decay tree. Binomial uncertainty are also shown.
    }
    \label{fig:heatmap_performance}
\end{figure}

The dependence on both the topological tree depth and the number of reconstructed particles is shown in \figref{fig:heatmap_performance}.

The same trend is observed when considering both tree depth and particle multiplicity: the perfect reconstruction fraction decreases with increasing complexity, while the clustering efficiency remains typically between \(85\%\) and \(90\%\). This further supports the conclusion that reconstructing the internal hierarchy is the main limitation. A more pronounced drop occurs for depth-1 trees with 8 particles; however, these topologies arise from the topologization procedure and appear only three times in the dataset.

\subsection{Performance on exclusive decay channels}

The performance of the same model trained on the inclusive sample is also evaluated on a set of exclusive decay channels, with test samples of \(10\,000\) events per channel. The selected decay modes are commonly used as benchmarks in \(b\)-physics studies and cover a broad range of signatures, including rare flavor-changing neutral current decays~\cite{aaij_test_2017,boletti_angular_2020}, channels used in lepton flavor studies~\cite{aaij_test_2023}, CP violation~\cite{aubert_observation_2007} and mixing measurements~\cite{aaij_measurement_2019}, and high-yield hadronic decays used for detector calibration and reconstruction performance studies. The results are summarized in \tabref{tab:exclusive_results}.

\begin{table*}[!tp]
\caption{
Performance on selected $b$-hadron decay channels.
BA denotes the balanced accuracy, WH the wrong hierarchy fraction
and \(\epsilon\) the clustering efficiency.
The uncertainties on BA are obtained by propagating binomial
uncertainties on the class recalls, assuming independent particle pairs.
The uncertainties on the reconstruction fractions and clustering
efficiency are binomial statistical uncertainties.
No WH cases are observed for the two-body channels.
}
\label{tab:exclusive_results}
\centering
\small
\setlength{\tabcolsep}{2pt}
\begin{tabular}{|l|c|c|c|c|}
\hline
\rowcolor{GrayLight!60}
\textbf{Decay channel} &
\textbf{BA [\%]} &
\textbf{Perfect [\%]} &
\textbf{WH [\%]} &
\textbf{\(\epsilon\) [\%]} \\
\hline

$B^{0} \rightarrow D^{+}[K^{-}\pi^{+}\pi^{+}]D^{-}[K^{+}\pi^{-}\pi^{-}]$
& $91.63 \pm 0.07$ & $81.4 \pm 0.5$ & $15.6 \pm 0.5$ & $97.1 \pm 0.2$ \\
\hline

\rowcolor{GrayLight!60}
$B^{0} \rightarrow K^{+}\pi^{-}$
& $99.1 \pm 0.1$ & $97.0 \pm 0.3$ & $0.0$ & $97.0 \pm 0.3$ \\
\hline

$B^{0} \rightarrow K^{*0}[K^{+}\pi^{-}]\mu^{+}\mu^{-}$
& $90.9 \pm 0.2$ & $77.4 \pm 0.8$ & $17.1 \pm 0.7$ & $94.5 \pm 0.4$ \\
\hline

\rowcolor{GrayLight!60}
$B^{+} \rightarrow \bar{D}^{0}[K^{+}\pi^{-}\pi^{+}\pi^{-}]\pi^{+}$
& $84.8 \pm 0.2$ & $94.1 \pm 0.4$ & $2.2 \pm 0.3$ & $96.3 \pm 0.3$ \\
\hline

$B^{+} \rightarrow \bar{D}^{0}[K^{+}\pi^{-}]\pi^{+}$
& $96.9 \pm 0.1$ & $88.9 \pm 0.6$ & $7.3 \pm 0.5$ & $96.2 \pm 0.3$ \\
\hline

\rowcolor{GrayLight!60}
$B^{+} \rightarrow D_s^{+}[K^{+}K^{-}\pi^{+}]\bar{D}^{0}[K^{+}\pi^{-}]$
& $93.4 \pm 0.1$ & $80.2 \pm 0.7$ & $15.8 \pm 0.6$ & $95.9 \pm 0.3$ \\
\hline

$B^{+} \rightarrow J/\psi[e^{+}e^{-}]K^{+}$
& $93.6 \pm 0.3$ & $81 \pm 1$ & $14.8 \pm 0.9$ & $95.3 \pm 0.6$ \\
\hline

\rowcolor{GrayLight!60}
$B^{+} \rightarrow J/\psi[\mu^{+}\mu^{-}]K^{+}$
& $94.0 \pm 0.2$ & $82.8 \pm 0.7$ & $12.2 \pm 0.6$ & $95.0 \pm 0.4$ \\
\hline

$B^{+} \rightarrow K^{+}K^{-}\pi^{+}$
& $89.0 \pm 0.2$ & $71.4 \pm 0.8$ & $24.0 \pm 0.8$ & $95.3 \pm 0.4$ \\
\hline

\rowcolor{GrayLight!60}
$B_s^{0} \rightarrow D_s^{-}[K^{+}K^{-}\pi^{-}]\pi^{+}$
& $95.6 \pm 0.1$ & $90.1 \pm 0.5$ & $5.6 \pm 0.4$ & $95.7 \pm 0.4$ \\
\hline

$B_s^{0} \rightarrow J/\psi[\mu^{+}\mu^{-}]\phi[K^{+}K^{-}]$
& $88.7 \pm 0.1$ & $66.3 \pm 0.6$ & $26.7 \pm 0.6$ & $93.0 \pm 0.3$ \\
\hline

\rowcolor{GrayLight!60}
$\Lambda_b \rightarrow \Lambda_c^{+}[pK^{-}\pi^{+}]\pi^{-}$
& $94.1 \pm 0.1$ & $83.5 \pm 0.7$ & $12.4 \pm 0.6$ & $95.9 \pm 0.3$ \\
\hline

$\Lambda_b \rightarrow pK^{-}$
& $99.1 \pm 0.1$ & $96.3 \pm 0.3$ & $0.0$ & $96.3 \pm 0.3$ \\
\hline

\rowcolor{GrayLight!60}
$\Lambda_b \rightarrow pK^{-}\mu^{+}\mu^{-}$
& $94.5 \pm 0.1$ & $86.4 \pm 0.6$ & $8.0 \pm 0.5$ & $94.4 \pm 0.4$ \\
\hline

\end{tabular}
\end{table*}

 The performance is systematically better than for the inclusive sample, with perfect reconstruction fractions above \(66\%\) and clustering efficiencies between \(93\%\) and \(97\%\) for all of the considered decays. This improved performance can largely be attributed to the lower topological complexity of these exclusive decay channels. All exclusive channels considered here have a topological depth of one or two, whereas approximately \(35\%\) of the decays in the inclusive test sample have a topological depth of three. As discussed in Section~\ref{sec:perf_incl}, the reconstruction of the exact hierarchy becomes more challenging as the tree depth increases.

The clustering efficiency remains consistently high across the different channels, showing only small variations. The largest variation is observed in the perfect reconstruction fraction. Simple two-body decays such as \(B^{0}\rightarrow K^{+}\pi^{-}\) and \(\Lambda_b\rightarrow pK^{-}\) reach a perfect reconstruction fraction of \(97\%\), while channels containing intermediate states exhibit larger wrong hierarchy contributions. Overall, the results obtained on exclusive decays confirm the behaviour observed with the inclusive sample, and demonstrate robust reconstruction performance across a broad range of relevant $b$-hadron decay topologies.

\FloatBarrier
\subsection{Timing studies}

\begin{figure}[!htbp]
    \centering
    \includegraphics[width=1\linewidth]{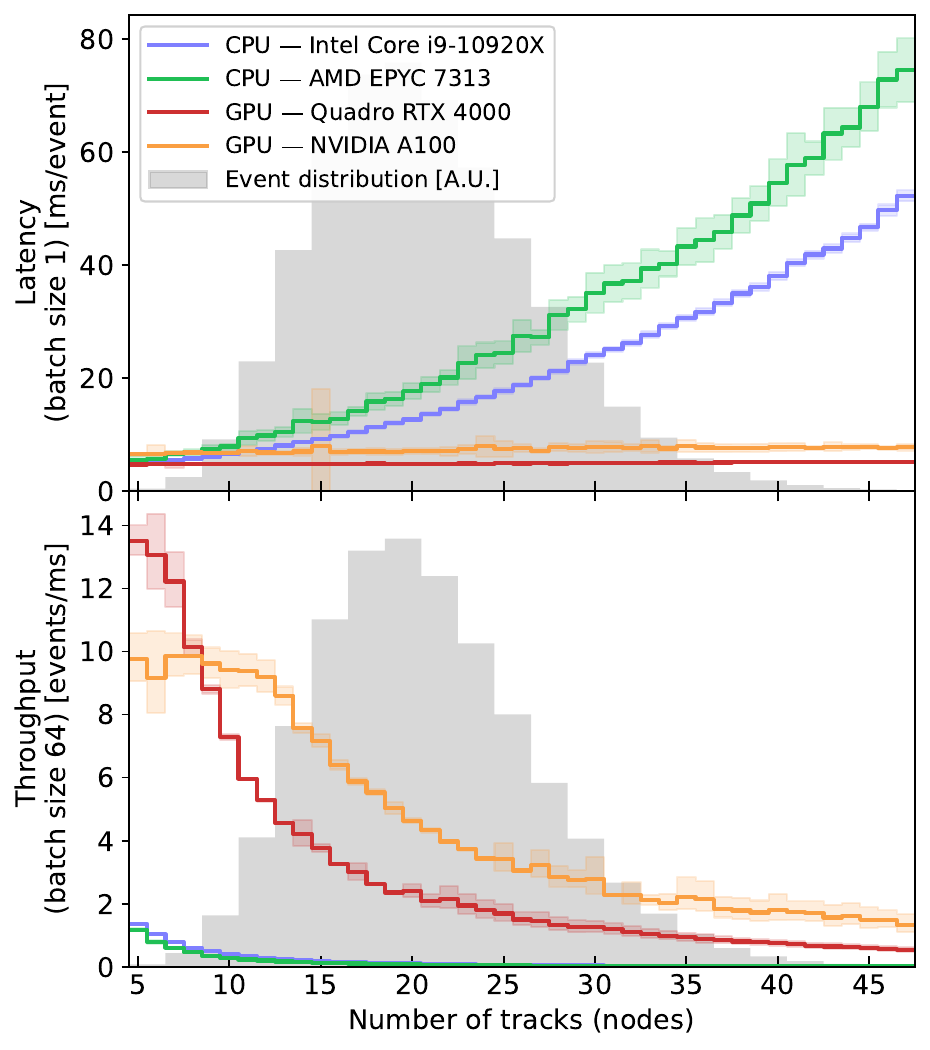}
\caption{
    Inference latency (top) and throughput for batches of 64 events
    (bottom) versus track multiplicity for two single-thread CPUs and
    two GPUs. Curves show means and colored bands indicate standard
    deviations. Grey histograms show the normalized track multiplicity distribution for easy visualization.
}
    \label{fig:timing}
\end{figure}

The inference latency and throughput are shown in \figref{fig:timing}
as functions of the track multiplicity, for two central processing units (CPUs) operated in single-thread mode, and for two graphics processing units (GPUs).

 Latency is measured using 100 individual events for each track multiplicity. The reported values and uncertainties are the mean and standard deviation of these measurements. CPU latency increases approximately quadratically with track
multiplicity, which is consistent with the $N(N-1)/2$ particle pairs in a fully
connected graph and the computational cost scaling linearly with the
number of edges. GPU latency remains approximately constant at
$5\,\mathrm{ms}$ for the NVIDIA Quadro RTX 4000 and $7\,\mathrm{ms}$ for
the NVIDIA A100, suggesting that fixed overheads dominate for the small
graphs considered here. A stronger dependence on graph size is expected
at higher multiplicities, beyond the studied range. Throughput is measured using batches of 64 events with equal track
multiplicity, each processed ten times. Values and uncertainties
represent the mean and standard deviation of the number of events
processed per unit time. Unlike latency, GPU throughput decreases with track multiplicity because processing 64 events at once makes GPU processing time more sensitive to graph size. Both GPUs outperform the single-thread
CPU configurations throughout the studied range.

\FloatBarrier
\section{Conclusions}

This work presents the first deep learning-based full event interpretation prototype for FCC-$ee$, using graph neural networks to identify hierarchical $b$-hadron decay trees starting from reconstructed charged particles. On the inclusive \(e^{+}e^{-}\rightarrow Z\rightarrow b\bar{b}\) test sample, the model achieves a perfect reconstruction fraction of approximately \(39\%\) and a clustering efficiency of about \(89\%\). The clustering efficiency remains high as the decay topology becomes more complex, showing that the model is able to reliably identify the correct particles belonging to a decay chain, while the main limitation remains the inference of the exact internal hierarchy of the decay tree. The benchmark exclusive channels show the same behaviour, with clustering efficiencies as high as \(93\%\) to \(97\%\) and perfect reconstruction fractions above \(66\%\). These results highlight the potential of graph-based event interpretation for heavy-flavor reconstruction at FCC-$ee$.

The present study constitutes a proof of concept. Future developments should extend the reconstruction to neutral particles and broaden the scope of the model to include charm-hadron decay chains. The inclusion of neutral particles could naturally benefit from heterogeneous graph neural networks, first explored in Ref.~\cite{sutcliffe_scalable_2025}, allowing different particles and relations to be explicitly represented in the graph \cite{elena_physics_2026}. Another important extension would be to move beyond LCAG and reconstruct richer decay tree information, including particle identities, as investigated by Belle II in Ref.~\cite{cerasoli_graph_2026}. Together, these developments could pave the way toward a more comprehensive event interpretation framework, capable of reconstructing a wider range of heavy-flavor decay topologies and providing a common foundation for precision flavor physics measurements at FCC-$ee$.

\subsection*{Acknowledgement}

The authors acknowledge colleagues from the LHCb and FCC collaborations for useful discussions. This work is supported by the SNSF Ambizione program with the grant n. PZ00P2\_202065, by the ERC program with the grant CLIMB 101077940, by the CNPq PQ program with the grant n. 310007/2026-7, and by the FAPERJ JCNE and Pensa Rio programs with the grants n. E-26/204.482/2025 and E-26/210.818/2025.

\subsection*{Data Availability Statement}
Data will be made available on reasonable request.

\subsection*{Code Availability Statement}
Code will be made available on
reasonable request.

\newcounter{mainfigurecount}
\setcounter{mainfigurecount}{\value{figure}}
\begin{appendices}
\setcounter{figure}{\value{mainfigurecount}}
\renewcommand{\thefigure}{\arabic{figure}}
\section{Hyperparameter optimization}
\label{app:hyperparameter_optimization}

\begin{figure}[!htbp]
    \centering
    \includegraphics[width=0.95\linewidth]{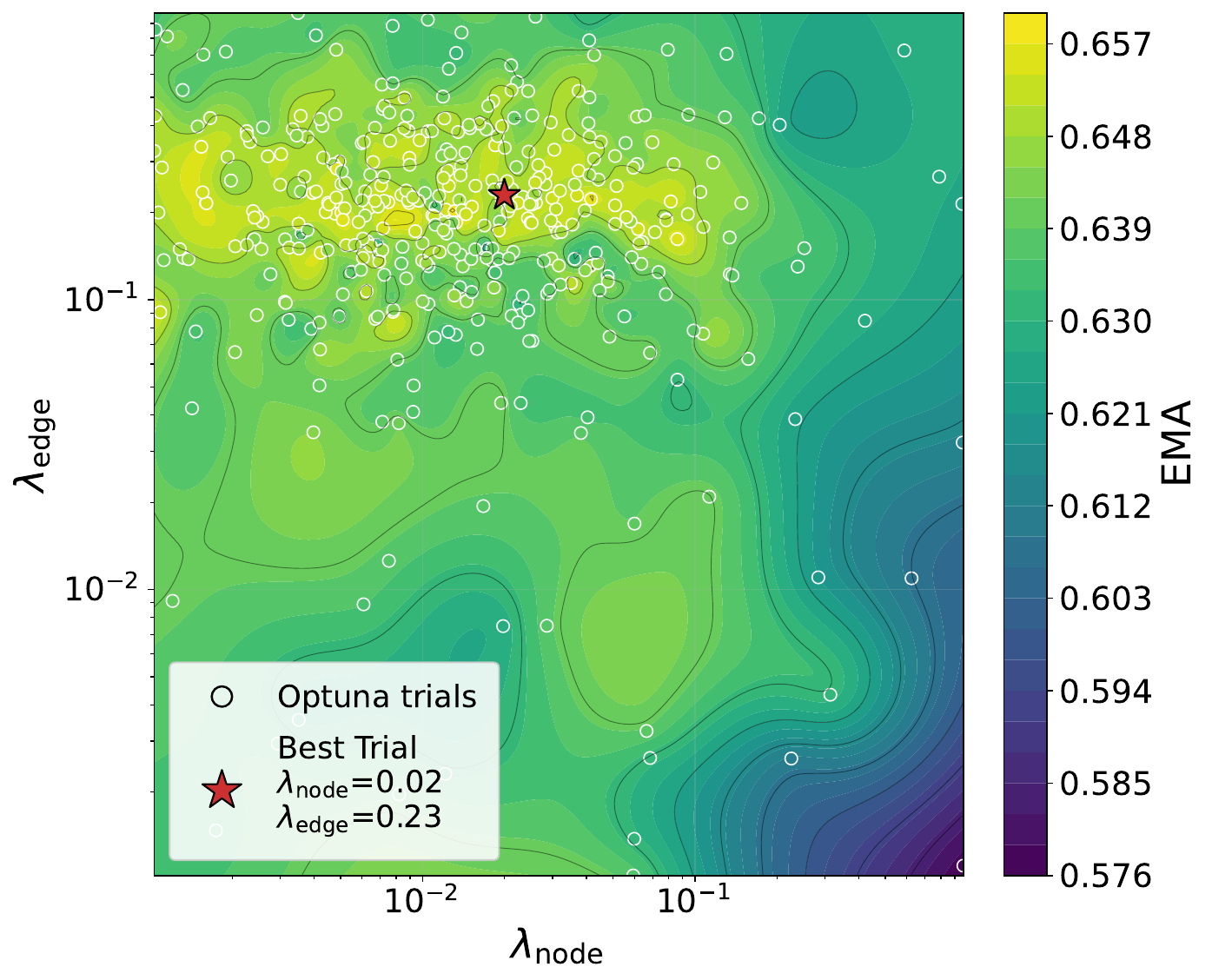}
    \caption{
    Optimization of the auxiliary loss coefficients \(\lambda_{\mathrm{node}}\) and \(\lambda_{\mathrm{edge}}\). The red star indicates the selected configuration, corresponding to \(\lambda_{\mathrm{node}}=0.02\) and \(\lambda_{\mathrm{edge}}=0.23\), while the circles denote the evaluated configurations.
    }
    \label{fig:opti_hp}
\end{figure}

The optimization of the auxiliary loss coefficients is shown in~\figref{fig:opti_hp}.

\end{appendices}

\bibliography{sn-bibliography}

\end{document}